\documentclass[lettersize,journal]{IEEEtran}
\usepackage{amsmath,amsfonts}
\usepackage{algorithmic}
\usepackage{algorithm}
\usepackage{array}
\usepackage[caption=false,font=normalsize,labelfont=sf,textfont=sf]{subfig}
\usepackage{textcomp}
\usepackage{stfloats}
\usepackage{url}
\usepackage{verbatim}
\usepackage{graphicx}
\usepackage{cite}
\usepackage{epstopdf}
\usepackage{multicol}
\usepackage{tabularx}
\usepackage{svg}
\usepackage{caption}
\usepackage{pifont}
\usepackage{hyperref} 
\usepackage{booktabs}
\usepackage{array}
\usepackage{threeparttable} 
\hypersetup{
    colorlinks=true,
    linkcolor=blue,
    citecolor=blue,
    urlcolor=blue
}
\usepackage{colortbl} 
\usepackage{xcolor} 
\usepackage{multirow}
\usepackage{amssymb}  
\begin{document}

\title{IDEAL: A Multimodal Domain Adaptation Framework for EEG-Eye Emotion Recognition}

\author{Yang Wu, Changde Du, He Chen, Tzyy-Ping Jung,~\IEEEmembership{Fellow,~IEEE}, Jinpeng Li,~\IEEEmembership{Senior Member,~IEEE}
\thanks{This work was supported in part by the National Key RD Program of China under Grant 2024YFB3312400, and in part by the National Natural Science Foundation of China under Grant 62106248. Corresponding author: Jinpeng Li (lijinpeng@scut.edu.cn).}
\thanks{Yang Wu, He Chen and Jinpeng Li are with the School of Automation Science and Engineering, South China University of Technology, Guangzhou 510641, Guangdong, China.}
\thanks{Changde Du is with State Key Laboratory of Brain Cognition and Brain-inspired Intelligence Technology, Institute of Automation, Chinese Academy of Sciences, Beijing 100190,
China.}
\thanks{Tzyy-Ping Jung is with the Department of Bioengineering, University of California at San Diego, La Jolla 92092, USA, and also with the Swartz Center for Computational Neuroscience, Institute for Neural Computation, University of California at San Diego, La Jolla 92093, USA.}
}

\markboth{Journal of \LaTeX\ Class Files,~Vol.~14, No.~11, August~2025}%
{Shell \MakeLowercase{\textit{et al.}}: A Sample Article Using IEEEtran.cls for IEEE Journals}


\maketitle

\begin{abstract}
Electroencephalography (EEG) emotion recognition serves as a pivotal interface for human-computer interaction, yet the physiological variability across individuals complicates the already challenging task of fusing heterogeneous physiological signals (e.g., EEG and eye movements). However, most prevalent domain adaptation paradigms are tailored for unimodal scenarios, failing to address the heterogeneity of multimodal signals. Furthermore, they predominantly rely on feature-level alignment, overlooking the fundamental data-level discrepancy, which risks compromising fine-grained discriminative information during aggressive adaptation. To bridge these coupled gaps, we propose Instance-based Domain Expansion and Adversarial Learning (IDEAL), a unified framework that synergizes \emph{instance-level curriculum expansion} with \emph{feature-level hierarchical adversarial alignment}. IDEAL first introduces a multi-model collaborative screening mechanism, which propagates high-confidence target samples to explicitly bridge the distributional gap at the data level via a quantity-quality equilibrium strategy. We provide a theoretical analysis that this instance expansion strategy strictly tightens the upper bound of the target risk. Subsequently, a hierarchical adversarial network, augmented with the angular-contrastive constraints, progressively aligns representations from low-level statistics to high-level semantics while preserving class separability. Extensive experiments on four benchmark datasets demonstrate that IDEAL significantly outperforms state-of-the-art methods. To facilitate reproducibility and future research, our source code is publicly available at \url{https://github.com/WY-BCI-Club/IDEAL}.
\end{abstract}


\section{Introduction}

\IEEEPARstart{E}{motion} recognition helps machines perceive, understand, and respond to human emotions. As a core technology driving human-computer interaction toward deeper and more intelligent development, it holds broad prospects in fields such as intelligent healthcare \cite{Saha2026ADD, MENG2025130418}, driver state monitoring \cite{11423591, Song2020EEG, XIAN2026117976}, and personalized services \cite{11613993}.

Among various physiological signals, EEG stands out as a pivotal modality for probing emotional processes due to its direct capture of cortical neural activity, high temporal resolution, and reliability \cite{Zheng2015Investigating}. EEG-based emotion recognition has evolved from early methods relying on handcrafted features with traditional classifiers to deep learning approaches~\cite{DGCNN}. Despite these advancements, the inherent low signal-to-noise ratio and non-stationarity of EEG signals fundamentally limit the accuracy and robustness of unimodal solutions \cite{CSMM}.

Multimodal emotion recognition has emerged as a viable solution to address these inherent limitations, leveraging complementary information from diverse signals such as EEG, facial expressions, eye movements, and electrocardiograms. Among these combinations, the fusion of EEG and eye movement signals stands out. For example, CFDA-CSF alleviates modal heterogeneity by aligning modalities from coarse to fine granularity~\cite{Jimenez2024CFDA}. Cheng et al. proposed a dense graph convolution network to integrate multimodal spatial topology and consistency \cite{DG-JCA}. Zhu et al. introduced a cross-attention mechanism to achieve a deep fusion of EEG and eye movement signals \cite{CSMM}. Despite these innovations, existing EEG-Eye fusion paradigms still suffer from critical drawbacks: \emph{they often exhibit inherent modality bias (e.g., rigid guidance strategies that over-rely on EEG) and superficial integration of cross-modal information, failing to adequately address the hierarchical feature alignment required to bridge the semantic gap between neural and behavioral signals.}

Beyond modal heterogeneity, the domain shift induced by individual differences poses a paramount challenge to the deployment of emotion recognition. Transfer learning for this task typically involves two approaches: feature-level and instance-level domain adaptation. Feature-level methods aims to learn domain-invariant representations. Ganin et al. pioneered domain-adversarial neural networks (DANN), which integrates a gradient reversal layer to align source-target feature distributions by confusing the discriminator \cite{Ganin2016Domain}. Another classic approach is maximum mean miscrepancy (MMD) to achieve feature alignment \cite{MMD}. Complementarily, instance-level transfer learning leverages high-quality source instances directly. Wang et al. proposed instance-based deep transfer learning, which screens source instances via pre-trained model evaluations \cite{wang2019instancebased}. Zhou et al. designed EEGMatch, a semi-supervised framework incorporating sample pairing to select reliable target samples, addressing incomplete labels \cite{EEGMatch}. Nevertheless, these methods face a dilemma: \emph{aggressive domain alignment in feature-level approaches risks compromising emotional feature discriminability (i.e., negative transfer), while instance-level methods struggle to balance the ``quantity-quality" trade-off of selected instances and remain highly sensitive to hyperparameter settings}.

To bridge the coupled gaps of multimodal heterogeneity and the conflict between feature alignment and discriminability, we propose a unified framework named \textbf{I}nstance-based \textbf{D}omain \textbf{E}xpansion and \textbf{A}dversarial \textbf{L}earning (\textbf{IDEAL}). It adopts a full-chain design that synergizes \emph{instance-level curriculum expansion} with \emph{feature-level hierarchical adaptation}. Specifically: (1) \textbf{Instance-level Bridging.} We mitigate instance selection bias through the multimodal collaborative screening. It quantifies sample reliability via prediction consistency and employs a dynamic quantity-quality equilibrium strategy to propagate high-confidence target samples, narrowing the raw distributional gap. (2) \textbf{Hierarchical Feature Interaction}. We resolve the disconnection in multimodal adaptation by constructing a hierarchical feature interaction module. This captures modality specificity through EEG temporal and eye spatial embeddings, allocates fusion weights via an adaptive Transformer, and employs a three-level adversarial discriminator to progressively align features from shallow statistics to deep semantics. (3) \textbf{Discriminability Enhancement}. We alleviate the optimization conflict between alignment and classification by incorporating ArcFace loss \cite{Deng2019ArcFace} to enhance intra-class compactness, contrastive loss to reinforce emotion-specific features, and a distillation-based regularization network to suppress overfitting to individual-specific noise. Our core contributions are summarized as follows:

(1) We propose IDEAL, a novel multimodal domain adaptation framework that  integrates instance-level curriculum expansion and feature-level transfer within a unified architecture. It addresses the coupled challenges of domain shift caused by individual differences and modal heterogeneity.

(2) We propose a multimodal curriculum expansion mechanism. By quantifying sample confidence via prediction consistency and a quantity-quality balance objective, it enables the selection of high-value target samples without target labels, effectively narrowing the distribution gap at the data level.

(3) We introduce a multimodal transformer to mine complementary information between EEG and eye movement signals, augmented by a hierarchical adversarial mechanism, to progressively align features at different levels.

(4) By enhancing feature discriminability through a combination of angular and contrastive objectives, IDEAL mitigates negative transfer and achieves state-of-the-art performance on cross-subject tasks across four public datasets.

\section{Related Work}

\subsection{Domain Adaptation}
Domain adaptation algorithms have evolved from early statistical feature alignment to contemporary deep adversarial adaptation. Early approaches primarily focused on minimizing distribution discrepancies in a projected feature space. For example, transfer component analysis (TCA) \cite{MMD} minimizes the MMD between domains. Extending this, joint distribution adaptation (JDA) \cite{JDA} simultaneously aligns marginal and conditional distributions by leveraging pseudo-labels to approximate target class posteriors.

With the advent of deep learning, DANN \cite{Ganin2016Domain} integrates a gradient reversal layer (GRL) and forces the feature extractor to learn domain-invariant representations that confuse the discriminator. Building on this, multi-adversarial domain adaptation (MADA) \cite{MADA} deployed class-specific discriminators to achieve fine-grained class-conditional alignment, effectively mitigating the boundary blurring inherent in global alignment strategies. Addressing the complexity of multi-subject data, multisource domain adversarial network (MDAN) \cite{MDAN} models the relationships between multiple source domains and the target. However, they predominantly treat all source samples equally during alignment. This oversight ignores the potential negative transfer caused by low-quality or outlier instances, leaving the data-level distributional gap unaddressed. \textit{Some surveys have systematically reaffirmed that distribution discrepancies between domains can result in negative transfer, while the presence of noisy negative samples in feature alignment further exacerbates this challenge \cite{contrastive2026survey}.}

\subsection{EEG Emotion Recognition}
EEG-based emotion recognition has witnessed a paradigm shift from handcrafted features to deep representation learning, with a specific focus on tackling cross-subject variability. Song et al. pioneered the dynamic graph convolutional neural network, which adaptively learns functional connectivity patterns to capture dynamic brain dependencies~\cite{DGCNN}. Zhang et al. developed the temporal adaptive sampling network, formulating the selection of salient EEG segments as a Markov decision process via deep reinforcement learning, thereby overcoming fixed-label dependencies in unsupervised scenarios \cite{TAS-Net}.

To further enhance cross-subject generalization, Shen et al. proposed the dynamic attention-based EEG state transition framework, which models parallel neural processes and their transitions across individuals \cite{DAEST}. Yang et al. introduced spectral-spatial attention alignment to learn emotion-specific cognitive attributes to overcome individual differences \cite{S2A2-MSDA}. Cui et al. leveraged gated recurrent units with minimum class confusion to refine the discrimination of subtle emotional states \cite{GRU-MCC}. Despite these architectures, unimodal EEG solutions remain constrained by the signal's inherent non-stationarity and low signal-to-noise ratio\cite{Yilmaz2026EEG}. \textit{ This performance plateau underscores the necessity of integrating complementary modalities, such as eye movements, to construct more robust recognition systems.}

\subsection{Multimodal Emotion Recognition}
Recent research has pivoted towards multimodal emotion recognition. Gong et al. proposed the coordinated representation decision fusion network, combining inter-modal correlation maximization with decision-level fusion to address information loss and noise sensitivity \cite{gong2023codf}. Tang et al. developed the robust heterogeneous physiological representation network, employing a hierarchical fusion module with scene-adaptive pre-training \cite{tang2024hierarchical}. Zhu et al. proposed multi-dimensional homogeneous encoding space alignment (MHESA), utilizing multi-task joint optimization to dynamically adjust spatial weights \cite{zhu2024eeg}. Jimenez et al. introduced multimodal multisource domain adaptation (MMDA), which combats negative transfer via prioritized distribution alignment and cross-modal feature selection \cite{10819285}. CSMM utilized momentum encoders and cross-attention mechanisms to align EEG and eye movement timestamps for deep fusion \cite{CSMM}. However, most existing methods predominantly focus on feature-level alignment, assuming that the data distribution is amenable to direct adaptation. They largely neglect instance-level transfer and struggles to simultaneously address the coupled challenges of \emph{modal heterogeneity} (how to fuse) and \emph{data-level domain shift} (what to transfer) in high-variability cross-subject scenarios.

\
\section{Methodology}
\begin{figure*}
    \centering
    \includegraphics[width=1\linewidth]{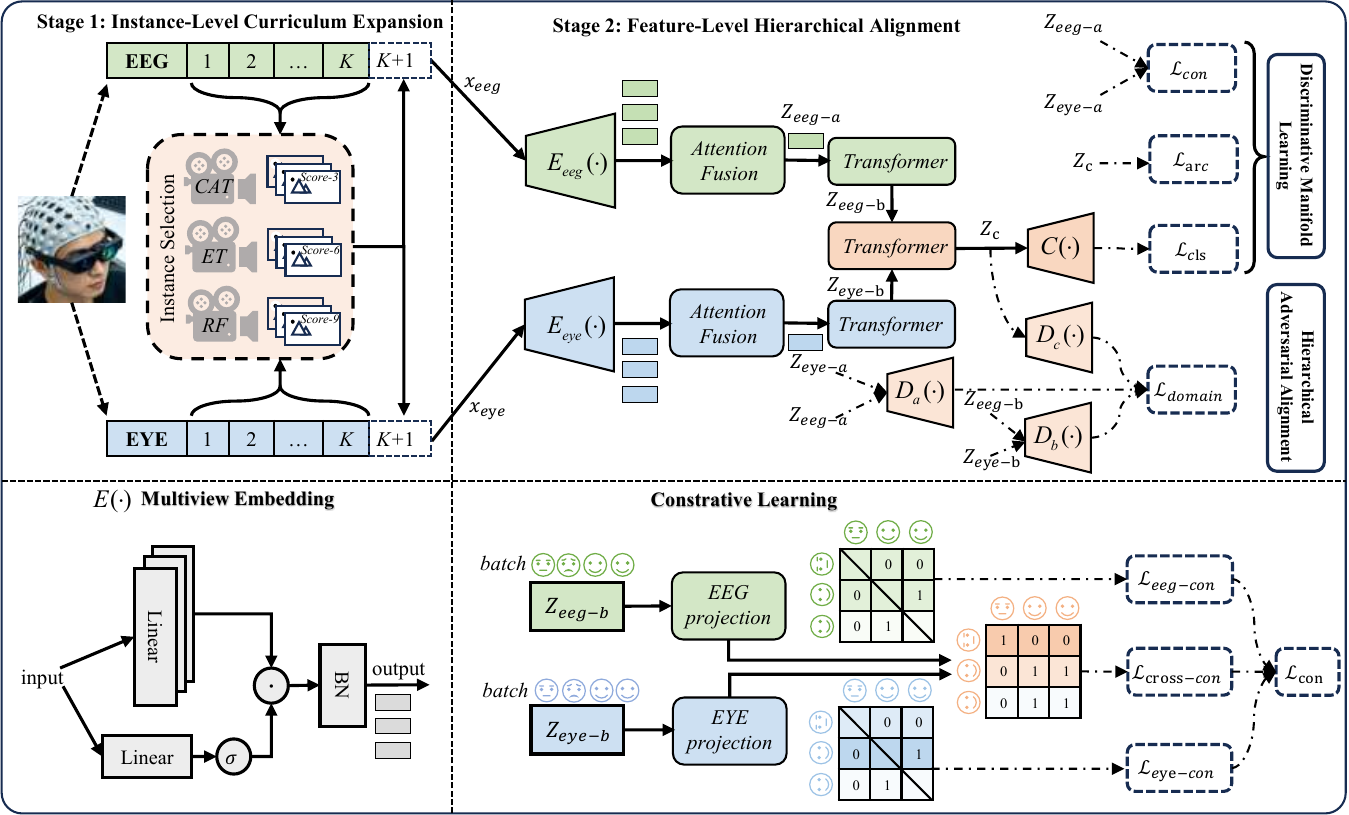}
    \caption{\small \textbf{The framework of Instance-based Domain Expansion and Adversarial Learning (IDEAL)}. It consists of two synergistic stages: Stage 1 (Instance-Level Domain Expansion) selects and augments high-confidence target domain samples via a collaborative screening mechanism to narrow source-target distribution gaps; Stage 2 (Feature-Level Hierarchical Adaptation) fuses EEG and eye movement features via Transformers and aligns them using a three-level adversarial strategy.}
     \label{figure:1}
\end{figure*}

\subsection{Problem Formulation, Modeling and Analysis}
This section systematically elaborates on the proposed IDEAL framework. We formulate the cross-subject emotion recognition task as a domain adaptation problem under the assumption of covariate shift.

Let $\mathcal{D}_S = \{(\mathbf{x}_i^S, y_i^S)\}_{i=1}^{N_S}$ denote the source domain with $N_S$ labeled samples collected from $K$ subjects, where $\mathbf{x}_i^S = [\mathbf{x}_{i, \text{eeg}}^S; \mathbf{x}_{i, \text{eye}}^S]$ represents the multimodal input and $y_i^S \in \{1, \dots, N_c\}$ is the corresponding emotion label. Similarly, let $\mathcal{D}_T = \{\mathbf{x}_j^T\}_{j=1}^{N_T}$ denote the target domain with $N_T$ unlabeled samples from a new subject. The marginal distributions of the two domains differ, i.e., $P_S(\mathbf{x}) \neq P_T(\mathbf{x})$, due to individual physiological variability.

\textbf{Assumption of Shared Latent Structure.} We posit the existence of a shared latent subspace that can be reliably estimated from samples of both the source and target domains. Within this subspace, we aim to identify a reliable subset $C \subseteq \mathcal{D}_T$ that captures the shared structure across domains. The marginal distribution $P_C$ of this subset is expected to satisfy two properties: (1) \textit{Affinity}: it should be distributionally close to the source domain ($P_C \approx P_S$), facilitating effective knowledge transfer; (2) \textit{Representativeness}: it should partially preserve the intrinsic information of the target domain ($P_C \approx P_T$), ensuring sufficient target relevance. 

Accordingly, we formulate the construction of $C$ as a constrained subset selection problem:
\begin{equation}
\max_{C \subseteq \mathcal{D}_T} |C| \quad \text{s.t.} \quad d(P_C, P_S) \leq \epsilon_1, \; d(P_C, P_T) \leq \epsilon_2,
\label{eq:subset_optimization}
\end{equation}
where $d(\cdot, \cdot)$ denotes a distributional divergence metric (e.g., MMD or $\mathcal{H}\Delta\mathcal{H}$-distance), $|C|$ is the cardinality of the selected subset, and $\epsilon_1, \epsilon_2 > 0$ are tolerance thresholds. This formulation seeks to maximize information transfer (\emph{quantity}) while bounding the negative transfer risk (\emph{quality}).

\textbf{Theoretical Justification.} To theoretically validate the necessity of constructing such a subset, we analyze the generalization error bound based on the theory proposed by Ben-David et al. \cite{ben2010theory}. Let $\mathcal{H}$ be a hypothesis space of VC-dimension $d$. For any hypothesis $h \in \mathcal{H}$, the expected error on the target domain $\epsilon_T(h)$ is bounded by:
\begin{equation}
\epsilon_T(h) \leq \epsilon_S(h) + \frac{1}{2} d_{\mathcal{H}\Delta\mathcal{H}}(\mathcal{D}_S, \mathcal{D}_T) + \lambda 
\end{equation}
where $\epsilon_S(h)$ is the empirical risk on the source domain, $d_{\mathcal{H}\Delta\mathcal{H}}$ is the divergence measuring the discrepancy between two distributions, and $\lambda$ is the ideal combined error.

\textbf{Proposition 1 (Strict Bound Tightening under Effective Screening).} 
Let $\mathcal{D}_S^* = \mathcal{D}_S \cup C$ be the augmented source domain, where $C \subseteq \mathcal{D}_T$ is non-empty. Under an effective screening mechanism that identifies a subset $C$ satisfying \textbf{Condition 1}:
\begin{equation}
d_{\mathcal{H}\Delta\mathcal{H}}(C, \mathcal{D}_T) < d_{\mathcal{H}\Delta\mathcal{H}}(\mathcal{D}_S, \mathcal{D}_T),
\label{eq:screening_condition}
\end{equation}
the domain divergence after augmentation is strictly reduced, i.e., $d_{\mathcal{H}\Delta\mathcal{H}}(\mathcal{D}_S^*, \mathcal{D}_T) < d_{\mathcal{H}\Delta\mathcal{H}}(\mathcal{D}_S, \mathcal{D}_T)$. Consequently, the target generalization upper bound is strictly tightened (with $C = \varnothing$ as the only degenerate case). This condition serves as the theoretical criterion for screening effectiveness, and our goal in Section~III-B is to design a mechanism fulfilling this proposition. Rigorous mathematical proof is provided in the Supplementary Material.

Guided by this theoretical insight, we formulate the unified transfer learning objective to define a deep neural network composed of a feature extractor $F$ and a classifier $G$. The overarching objective is to learn a hypothesis $G \circ F$ that minimizes the expected risk on $\mathcal{D}_T$ by leveraging the augmented source domain $\mathcal{D}_S \cup C$:

\begin{equation}
\min_{F, G}\left\{\mathbb{E}_{(\mathbf{x}, y) \sim \mathcal{D}_S \cup C} \mathcal{L}_{\text{cls}}(G(F(\mathbf{x})), y) + \lambda\mathcal{L}_{\text{align}}(F)\right\}
\label{eq:total_objective}
\end{equation}

where $\mathcal{L}_{\text{cls}}$ is the cross-entropy classification loss, $\mathcal{L}_{\text{align}}$ represents the domain alignment loss (e.g., adversarial or contrastive constraints), and $\lambda$ is trade-off hyperparameters.

Figure \ref{figure:1} illustrates the IDEAL framework. The training process proceeds in two synergistic stages. \textbf{Stage 1 (Instance-Level Domain Expansion):} A multi-model collaborative mechanism screens high-confidence target samples to construct an intermediate domain $\mathcal{D}_{int}$, explicitly narrowing the raw data-level gap as theoretically justified above. \textbf{Stage 2 (Feature-Level Hierarchical Adaptation):} The expanded data feeds into a multimodal transformer network, where a hierarchical adversarial strategy aligns representations from low-level statistics to high-level semantics.

\subsection{Instance-Level Domain Expansion}
\label{sec:domain_expansion}

 Building upon Proposition 1, we now instantiate a practical mechanism to identify the subset \(C\) that satisfies condition (\ref{eq:screening_condition}). Recall that this condition requires \(d_{\mathcal{H}\Delta\mathcal{H}}(C, \mathcal{D}_T) < d_{\mathcal{H}\Delta\mathcal{H}}(\mathcal{D}_S, \mathcal{D}_T)\). Direct evaluation is infeasible due to the unknown target distribution \(P_T\). We therefore reformulate screening as \textit{observer-based estimation}: consensus among multiple source-trained observers serves as a proxy for a sample's proximity to the shared subspace.

\textit{1) Consensus of Multiple Observers.} We employ three heterogeneous observers (Random Forest \cite{breiman2001random}, Extremely Randomized Trees \cite{geurts2006extremely}, and CatBoost \cite{CAT}). Each observer is trained with three input views: EEG-only, eye-only, and their concatenation, with aggregation weights \(\{0.3, 0.3, 0.4\}\). For target sample \(\mathbf{x}_j\), the aggregated prediction is:
\begin{equation}
P_{\mathrm{agg}}(\mathbf{x}_j) = \frac{1}{3} \sum_{m=1}^{3} p_m(\mathbf{x}_j),
\label{eq:aggregated}
\end{equation}
where \(p_m(\mathbf{x}_j)\) is the output of the \(m\)-th observer. The \textit{consensus score} is defined as:
\begin{equation}
\mathrm{Cons}(\mathbf{x}_j) = \max_k [P_{\mathrm{agg}}(\mathbf{x}_j)]_k
\label{eq:consensus}
\end{equation}
with pseudo-label \(\hat{y}_j = \arg\max_k [P_{\mathrm{agg}}(\mathbf{x}_j)]_k\). High consensus indicates observer agreement, implying the sample resides in a source-covered region of the shared subspace. Such samples yield low observer-based divergence estimation error, thus probably aligning with  \textbf{Condition 1}.

\textit{2) \textit{Subset Optimization.}} Samples are sorted by descending consensus. Let \(r(t) \in [0,1]\) denote the proportion selected by threshold \(t\), i.e., \(C_t = \{\mathbf{x}_j \mid \mathrm{Cons}(\mathbf{x}_j) \geq t\}\). The selection ratio \(r\) involves a critical trade-off: conservative selection (small \(r\)) ensures quality but limited information gain, while aggressive selection (large \(r\)) risks introducing low-consensus samples that violate \textbf{Proposition 1} and cause negative transfer. We formalize this as:
\begin{equation}
\max_{t} \mathcal{J}(t) = \alpha \cdot \mathrm{Cons}_{\min}(t) + (1-\alpha) \cdot \bigl(r(t) - r(t)^2\bigr)
\label{eq:optimization}
\end{equation}
where \(\mathrm{Cons}_{\min}(t) = \min_{\mathbf{x}_j \in C_t} \mathrm{Cons}(\mathbf{x}_j)\) serves as a conservative quality measure. The term \(r(t) - r(t)^2\) penalizes excessive quantity, reflecting the principle that quality is prioritized over quantity—low-quality samples incur a far more severe cost due to negative transfer than the benefit gained from additional samples.

The trade-off parameter \(\alpha\) adapts to domain shift:
\begin{equation}
\alpha = 1 - u, \quad u = \frac{1}{N_T}\sum_{j=1}^{N_T} \mathrm{Cons}(\mathbf{x}_j)
\label{eq:alpha}
\end{equation}
Here $u$ quantifies the overall quality of target samples as perceived from the source domain, reflecting cross-domain overlap. Guided by our theory, low $u$ (severe domain shift) increases $\alpha$ to prioritize quality, while high $u$ decreases $\alpha$ to favor quantity. This adaptive design avoids manual threshold tuning, which is notoriously dataset- and subject-sensitive.

\textit{3) Solution.} We solve Eq. (\ref{eq:optimization}) via a simple traversal search over \(t\).  The final expanded source domain is \(\mathcal{D}_S^* = \mathcal{D}_S \cup C_t\), which, according to\textbf{ Proposition 1}, yields a strictly tightened target generalization bound.
`

\subsection{Network Framework}
\label{sec:network_framework}

To achieve robust cross-subject emotion recognition, we design a hierarchical feature extraction pipeline that progressively elevates representations from low-level signals to high-level semantic embeddings. As illustrated in Figure \ref{figure:1} (Stage 2), the framework comprises three logical blocks: Multi-view Embedding for feature disentanglement, Hierarchical Transformer for intra- and inter-modal interaction, and a Decision Head for emotion prediction.

\subsubsection{Multi-view Embedding and Attention-Fusion}

Drawing inspiration from \cite{Jiang2024SEED}, we employ a multi-view embedding module to decompose physiological signals into diverse semantic subspaces. 
Let $\mathbf{x} \in \mathbb{R}^{B \times D_{\text{in}}}$ denote the input batch for a specific modality (EEG or Eye), where $B$ is the batch size. We generate $V$ parallel view representations via independent linear projections:
\begin{equation}
\mathbf{e}_j = \mathbf{x}\mathbf{W}_j + \mathbf{b}_j, \quad j = 1, 2, \dots, V
\label{eq:embedding_proj}
\end{equation}
where $\mathbf{e}_j \in \mathbb{R}^{B \times D_{\text{emb}}}$ is the $j$-th view embedding, and $\mathbf{W}_j, \mathbf{b}_j$ are learnable parameters initialized independently to encourage semantic diversity\cite{Jiang2024SEED}. 

To dynamically calibrate the contribution of different views, we employ a global gating vector $\mathbf{g} \in \mathbb{R}^{B \times V}$ to quantify the informativeness of each view:
\begin{equation}
\mathbf{g} = \sigma(\mathbf{x}\mathbf{W}_{\text{gate}} + \mathbf{b}_{\text{gate}})
\label{eq:gating_vector}
\end{equation}
where $\sigma(\cdot)$ is the Sigmoid activation function scaling weights to $[0, 1]$. We then obtain the weighted view embedding $\hat{\mathbf{e}}_j$ via element-wise multiplication:
\begin{equation}
\hat{\mathbf{e}}_j = \mathbf{e}_j \odot \mathbf{g}_j
\label{eq:weighted_embedding}
\end{equation}
This operation suppresses irrelevant noise while enhancing affect-salient features. The $V$ weighted embeddings are stacked to form a tensor $\mathbf{Z} \in \mathbb{R}^{B \times V \times D_{\text{emb}}}$, followed by batch normalization to mitigate internal covariate shift.

Consequently, we obtain the initial multi-view feature banks for EEG and eye movements, denoted as $\mathbf{Z}_{\text{eeg}}$ and $\mathbf{Z}_{\text{eye}}$, respectively. These features are subsequently aggregated into a compact representation via an attention fusion module (Figure \ref{figure:1}). Taking EEG as an example, we compute a global importance score $\boldsymbol{\alpha}_{\text{eeg}}$ using a learnable attention vector $\mathbf{w}_{\text{attn}}$:
\begin{equation}
\boldsymbol{\alpha}_{\text{eeg}} = \text{Softmax}(\mathbf{Z}_{\text{eeg}} \cdot \mathbf{w}_{\text{attn}})
\label{eq:attention_weights}
\end{equation}
The aggregated abstract feature $\mathbf{Z}_{\text{eeg-a}}$ is derived by the weighted summation:
\begin{equation}
\mathbf{Z}_{\text{eeg-a}} = \sum_{v=1}^{V} \boldsymbol{\alpha}_{\text{eeg}, v} \cdot \mathbf{Z}_{\text{eeg}, v}
\label{eq:fusion_sum}
\end{equation}
Similarly, we obtain eye movement feature $\mathbf{Z}_{\text{eye-a}}$.

\subsubsection{Hierarchical Feature Interaction}
To model the complex dependencies within and between modalities, we propose a two-stage Transformer architecture comprising Intra-modal Refinement and Cross-modal Synergy.

\textbf{1) Intra-modal Refinement:}
The aggregated features $\mathbf{Z}_{\text{eeg-a}}$ and $\mathbf{Z}_{\text{eye-a}}$ are first processed by independent Unimodal Transformers to model long-range temporal or spectral dependencies.We employ multi-head self-attention (MHSA). Given a query $\mathbf{Q}$, key $\mathbf{K}$, and value $\mathbf{V}$ derived from the input $\mathbf{X}$, the attention output for the $i$-th head is computed as:
\begin{equation}
\text{Head}_i(\mathbf{X}) = \text{Softmax}\left( \frac{(\mathbf{X}\mathbf{W}_i^Q) (\mathbf{X}\mathbf{W}_i^K)^T}{\sqrt{d_k}} \right) (\mathbf{X}\mathbf{W}_i^V)
\label{eq:scaled_dot_product}
\end{equation}
We concatenate and project the outputs from all heads. A feed-forward network (FFN) the processes these outputs using residual connections and layer normalization. This stage yields refined modality-specific features $\mathbf{Z}_{\text{eeg-b}}$ and $\mathbf{Z}_{\text{eye-b}}$.

\textbf{2) Cross-modal Synergy:}
To bridge the semantic gap between neural and behavioral signals, we introduce a Mix-Transformer. It accepts the concatenation of the refined features $[\mathbf{Z}_{\text{eeg-b}}; \mathbf{Z}_{\text{eye-b}}]$ as input. Unlike simple summation, the Mix-Transformer dynamically models the joint distribution of EEG and eye movements, effectively mining complementary information. The final output is a deeply fused multimodal representation $\mathbf{Z}_{\text{c}}$, which encapsulates both modality-invariant and modality-specific affective cues.

\subsubsection{Emotion Prediction}
We feed the fused representation into a classifier $G$ (typically a linear projection) to generate the probability distribution over $N_c$ emotion categories:
\begin{equation}
\hat{\mathbf{y}} = G(\mathbf{Z}_{\text{c}})
\label{eq:prediction}
\end{equation}
The model is optimized using the Cross-Entropy loss:
\begin{equation}
\mathcal{L}_{\text{cls}} = -\frac{1}{B} \sum_{i=1}^{B} \sum_{c=1}^{N_c} y_{i,c} \log(\hat{y}_{i,c})
\label{eq:ce_loss}
\end{equation}
where $y_{i,c}$ is the binary indicator (0 or 1) if class $c$ is the correct classification for observation $i$, and $\hat{y}_{i,c}$ is the predicted probability.

To equip the IDEAL framework with robust discriminability and domain invariance, we construct a compound objective function derived from two strategic perspectives: \emph{Discriminative Manifold Learning} and \emph{Hierarchical Adversarial Alignment}.

\subsubsection{Discriminative Manifold Learning ($\mathcal{L}_{\text{cs}}$)}
We aim to learn a feature manifold where intra-class compactness and inter-class separability are simultaneously maximized. This is achieved by imposing constraints at both the feature level (via contrastive learning) and the decision boundary level (via angular margin loss). The combined Class Separation loss is formulated as:
\begin{equation}
\mathcal{L}_{\text{cs}} = \mathcal{L}_{\text{contrast}} + \mathcal{L}_{\text{arc}}
\label{eq:loss_cs}
\end{equation}

\begin{table*}[h]
\centering
\caption{Comparison of Intra-session Experimental Results (Accuracy \% $\pm$ Std)}
\label{tab:intra_session}
\renewcommand{\arraystretch}{1.25}
\setlength{\tabcolsep}{1.5pt}
\resizebox{\textwidth}{!}{%
\begin{tabular}{lcccccccccccc}
\toprule
\multirow{2}{*}{\textbf{Method}} & \multicolumn{4}{c}{\textbf{SEED}} & \multicolumn{4}{c}{\textbf{SEED-IV}} & \multicolumn{4}{c}{\textbf{SEED-V}} \\
\cmidrule(lr){2-5} \cmidrule(lr){6-9} \cmidrule(lr){10-13}
 & S1 & S2 & S3 & \textbf{Avg} & S1 & S2 & S3 & \textbf{Avg} & S1 & S2 & S3 & \textbf{Avg} \\
\midrule
DCCA \cite{Li2020EEG} & 87.60$\pm$9.68 & 80.48$\pm$12.54 & 90.22$\pm$6.62 & 85.92$\pm$9.61 & 54.71$\pm$11.34 & 64.17$\pm$10.09 & 63.80$\pm$11.29 & 60.89$\pm$10.91 & 49.88$\pm$8.57 & 52.18$\pm$17.18 & 52.64$\pm$18.57 & 51.57$\pm$14.77 \\
DANN \cite{Ganin2016Domain} & 87.11$\pm$7.76 & 81.66$\pm$15.06 & 88.37$\pm$10.44 & 85.71$\pm$11.09 & 76.43$\pm$11.29 & 82.51$\pm$9.12 & 79.45$\pm$9.18 & 79.46$\pm$9.86 & 74.10$\pm$14.45 & 81.80$\pm$16.84 & 80.68$\pm$16.83 & 78.86$\pm$16.04 \\
JDA \cite{Li2020Domain} & 87.03$\pm$6.88 & 85.02$\pm$11.97 & 87.56$\pm$8.48 & 86.54$\pm$9.11 & 76.53$\pm$10.05 & 81.83$\pm$9.17 & 79.73$\pm$9.13 & 79.36$\pm$9.45 & 73.88$\pm$15.73 & 82.23$\pm$14.79 & 81.94$\pm$18.51 & 79.35$\pm$16.40 \\
MWACN \cite{Zhu2022Multisource} & 89.67$\pm$8.55 & 83.71$\pm$18.05 & 89.13$\pm$9.29 & 87.50$\pm$11.96 & 72.38$\pm$9.76 & 76.61$\pm$7.96 & 72.67$\pm$12.37 & 73.89$\pm$10.03 & 75.82$\pm$16.79 & 84.10$\pm$15.78 & 85.59$\pm$16.13 & 81.84$\pm$16.23 \\
\midrule
CFDA-CSF \cite{Jimenez2024CFDA} & 93.05$\pm$6.22 & 85.87$\pm$13.52 & 91.20$\pm$8.39 & 90.04$\pm$9.38 & 85.72$\pm$11.02 & 89.60$\pm$6.65 & 86.88$\pm$10.79 & 87.40$\pm$9.49 & 88.49$\pm$12.32 & 91.37$\pm$12.92 & 91.57$\pm$15.49 & 90.48$\pm$13.58 \\
MACDB \cite{MACDB}& 88.16$\pm$11.02 & 83.21$\pm$12.43 & 88.68$\pm$9.75 & 86.68$\pm$11.07 & 82.38$\pm$9.81 & 86.62$\pm$6.52 & 86.10$\pm$10.02 & 85.03$\pm$8.78 & 85.98$\pm$12.06 & 85.65$\pm$16.01 & 87.82$\pm$10.45 & 86.48$\pm$12.84 \\
MMDA \cite{10819285} & 92.04$\pm$10.72 & 88.03$\pm$13.88 & 92.55$\pm$6.35 & 90.87$\pm$10.32 & 86.76$\pm$12.21 & 87.57$\pm$6.92 & 87.66$\pm$9.41 & 87.33$\pm$9.51 & 91.03$\pm$12.94 & 92.17$\pm$9.66 & 90.60$\pm$13.87 & 91.27$\pm$12.16 \\
CMSLNet \cite{chen2024comprehensive} & - & - & - & - & 78.39 & 86.13 & 85.23 & 83.25 & - & - & - & 87.32$\pm$14.81 \\
MHESA \cite{zhu2024eeg} & - & - & - & - & - & - & - & 83.15$\pm$9.84 & - & - & - & 87.32$\pm$14.81 \\
CSMM \cite{10938180} & - & - & - & 94.96$\pm$5.27 & - & - & - & 89.82$\pm$6.22 & - & - & - & 89.22$\pm$9.59 \\
CMGNN \cite{10570465} & - & - & - & - & - & - & - & 90.21$\pm$3.49 & - & - & - & - \\
\midrule
\rowcolor{gray!20}
\textbf{IDEAL} & \textbf{97.91$\pm$3.90} & \textbf{92.66$\pm$6.89}& \textbf{96.51$\pm$4.84} & \textbf{95.69$\pm$5.21} & \textbf{92.66$\pm$6.89} & \textbf{94.27$\pm$5.33}& \textbf{92.56$\pm$8.90}& \textbf{93.16$\pm$7.04} & \textbf{94.03$\pm$7.55} & \textbf{95.96$\pm$6.38} & \textbf{94.25$\pm$11.45} & \textbf{94.75$\pm$8.46} \\
\bottomrule
\end{tabular}%
}
\end{table*}
\textbf{1) Multi-perspective Contrastive Loss ($\mathcal{L}_{\text{contrast}}$):}
As illustrated in Figure \ref{figure:1}, we enforce semantic consistency through intra-modal and cross-modal contrastive constraints.

\textit{Intra-modal Contrast (InfoNCE):} Let $\mathbf{Z} \in \mathbb{R}^{B \times D}$ be the normalized feature batch for a given modality (e.g., EEG). We construct a similarity matrix $\mathbf{S} = \mathbf{Z}\mathbf{Z}^T / \tau$, where $\tau$ is a temperature scalar. For each sample $i$, the positive set $\mathcal{P}(i)$ contains indices of samples from the same class, excluding $i$ itself. We define the supervised InfoNCE loss as:
\begin{equation}
\mathcal{L}_{\text{intra}}(\mathbf{Z}) = -\frac{1}{B} \sum_{i=1}^B \frac{1}{|\mathcal{P}(i)|} \sum_{p \in \mathcal{P}(i)} \log \frac{\exp(\mathbf{S}_{i,p})}{\sum_{k \neq i} \exp(\mathbf{S}_{i,k})}
\label{eq:infonce}
\end{equation}

\textit{Cross-modal Consistency:} To align the semantic spaces of EEG ($\mathbf{Z}_{\text{eeg}}$) and Eye ($\mathbf{Z}_{\text{eye}}$), we minimize the discrepancy between their pairwise similarity patterns. We define the cross-modal alignment loss using binary cross-entropy on the affinity matrix $\mathbf{S}_{\text{cross}} = \mathbf{Z}_{\text{eeg}}\mathbf{Z}_{\text{eye}}^T$:
\begin{equation}
\begin{aligned}
\mathcal{L}_{\text{cross}} = -\frac{1}{B^2} \sum_{i,j} \Big[ & \mathbb{I}(y_i=y_j) \cdot \log(\sigma(\mathbf{S}_{\text{cross}}[i,j])) \\
& + \mathbb{I}(y_i \neq y_j) \cdot \log(1 - \sigma(\mathbf{S}_{\text{cross}}[i,j])) \Big]
\end{aligned}
\label{eq:cross_loss}
\end{equation}
where $\mathbb{I}(\cdot)$ is the indicator function. The total contrastive loss balances modality-specific structure and cross-modal alignment:
\begin{equation}
\mathcal{L}_{\text{contrast}} = \gamma \left( \mathcal{L}_{\text{intra}}(\mathbf{Z}_{\text{eeg}}) + \mathcal{L}_{\text{intra}}(\mathbf{Z}_{\text{eye}}) \right) + (1 - 2\gamma) \mathcal{L}_{\text{cross}}
\label{eq:total_contrast}
\end{equation}
and we set \(\gamma = 0.3\) to achieve balanced optimization.

\textbf{2) Angular Margin Loss ($\mathcal{L}_{\text{arc}}$):}
To further refine the decision boundary, we employ ArcFace loss on the final fused representation $\mathbf{z}_c$. Let $\theta_{i,y_i}$ be the angle between feature $\mathbf{z}_i$ and the weight vector $\mathbf{W}_{y_i}$ of the ground-truth class. We introduce an additive angular margin $m$ to enforce a stricter classification criterion:
\begin{equation}
\mathcal{L}_{\text{arc}} = -\frac{1}{B} \sum_{i=1}^{B} \log \frac{e^{s \cdot \cos(\theta_{i,y_i} + m)}}{e^{s \cdot \cos(\theta_{i,y_i} + m)} + \sum_{k \neq y_i} e^{s \cdot \cos\theta_{i,k}}}
\label{eq:arcface}
\end{equation}
where $s$ is the scaling factor. This forces the model to learn features that are angularly separable with a geodesic distance margin.

\subsubsection{Hierarchical Adversarial Alignment ($\mathcal{L}_{\text{domain}}$)}
To mitigate domain shift across different levels of abstraction, we implement a multi-scale adversarial strategy. We deploy three domain discriminators $\{D_l\}_{l \in \{a,b,c\}}$ corresponding to low-level (shallow), mid-level (intermediate), and high-level (deep) features. 
Let $\mathbf{h}_l$ denote the concatenated feature vector at level $l$ (e.g., $\mathbf{h}_a = [\mathbf{Z}_{\text{eeg-a}}; \mathbf{Z}_{\text{eye-a}}]$). Each discriminator attempts to classify the domain label $d_i \in \{0, 1\}$ (source/target), while the feature extractor minimizes this classification accuracy via gradient reversal. The hierarchical adversarial loss is:
\begin{equation}
\mathcal{L}_{\text{domain}} = \sum_{l \in \{a,b,c\}} \eta_l \cdot \mathbb{E}_{\mathbf{x}} \left[ - \log(D_l(\mathbf{h}_l)) \right]
\label{eq:adv_loss}
\end{equation}
\(\eta_{a,b,c}=\{0.2,0.3,0.5\}\)..This multi-granularity alignment ensures that domain-invariant patterns are captured from raw signal statistics up to semantic abstractions.

\textbf{Total Objective:}
The final optimization objective integrates the classification task with the aforementioned constraints:
\begin{equation}
\mathcal{J} = \mathcal{L}_{\text{cls}} + \lambda_1 \mathcal{L}_{\text{cs}} + \lambda_2 \mathcal{L}_{\text{domain}}
\label{eq:final_objective}
\end{equation}
and we set $\lambda_{1,2}=\{0.3,0.5\}$ are hyperparameters balancing the trade-off between \emph{discrimination} and \emph{alignment}.

\section{Experiments and Results}
\label{sec:experiments}

\subsection{Experimental Setup}

\paragraph{Datasets}
We evaluate IDEAL on four public multimodal emotion datasets: SEED, SEED-IV, SEED-V, and SEED-VII \cite{Zheng2015Investigating,Zheng2019Identifying,liu2022identifying,Jiang2024SEED}. Each dataset comprises EEG and eye movement signals with varying subject scales and emotion categories (3, 4, 5, and 7 classes, respectively). All features are Z-score normalized per subject per session.

\paragraph{Experimental Scheme}
We adopt two leave-one-subject-out (LOSO) partitioning strategies: (1) \textit{Intra-Session LOSO}: within each session, one subject serves as the target and the remaining subjects from the same session as the source; (2) \textit{All-Session LOSO}: all sessions of a subject are aggregated as the target, with all other subjects as the source. Performance is reported as mean accuracy with standard deviation.

\paragraph{Training Details}
The training consists of two stages. In the first stage, the screening mechanism selects $M$ high-confidence target samples. In the second stage, we randomly sample $N = \min(\max(1500, N_S\cdot M), 3000)$ source instances from the source domain, where $N_S$ denotes the total number of source samples, and combine them with the $M$ screened target samples for joint training.

\paragraph{Baselines}
We benchmark IDEAL against two categories of methods:
(1) \textbf{Classic Transfer Learning:} DANN \cite{Ganin2016Domain}, JDA \cite{Li2020Domain}, and MWACN \cite{Zhu2022Multisource}.
(2) \textbf{State-of-the-Art Multimodal Methods:} CFDA-CSF \cite{Jimenez2024CFDA}, MMDA \cite{10819285}, CMSLNet \cite{chen2024comprehensive}, MHESA \cite{zhu2024eeg}, CSMM \cite{10938180}, CMGNN \cite{10570465}, DCCA-AM \cite{liu2022identifying}, MACDB\cite{MACDB}, CoDF-Net \cite{gong2023codf}, MSBLS \cite{gong2024cross}, and RHPRNet \cite{tang2024hierarchical}.
All methods are evaluated under identical partitioning protocols to ensure fair comparison.

\subsection{Emotion Classification Performance}

\textbf{1) Intra-Session Results.} Table \ref{tab:intra_session} reports the comparison under the Intra-Session protocol. IDEAL achieves superior performance across all datasets. On SEED, we achieve an average accuracy of \textbf{95.69\%}, surpassing the runner-up CSMM (94.96\%). Notably, in the first session, accuracy reaches 97.91\%, outperforming CFDA-CSF by 4.86\%. On \textbf{SEED-IV} and \textbf{SEED-V}, IDEAL significantly outperforms all baselines. Importantly, our method demonstrates the lowest standard deviation on SEED (5.21\%) and SEED-V (8.46\%), compared to MMDA and CFDA-CSF with variances exceeding 10\%. This indicates that IDEAL exhibits strong stability against individual fluctuations.

\begin{figure*}[t]
    \centering
    \includegraphics[width=1.0\linewidth]{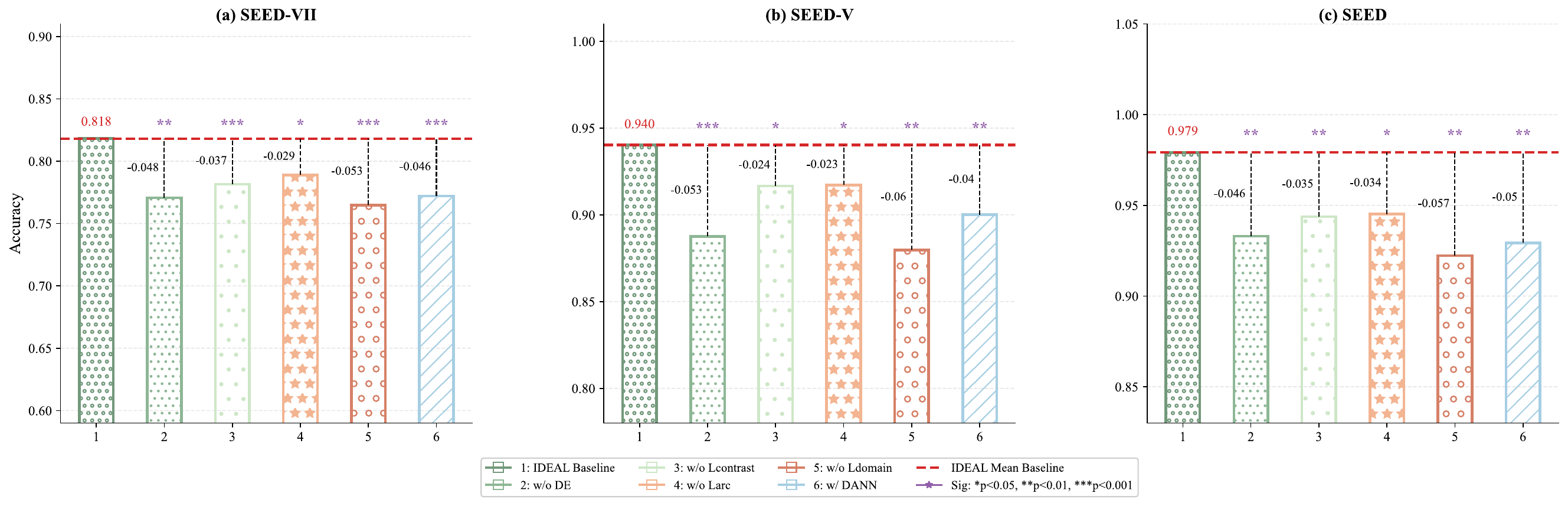}
    \caption{\small \textbf{Ablation Analysis}. Performance impact of removing key components across SEED, SEED-V, and SEED-VII datasets. Error bars indicate standard deviation. Significance levels: *** $p<0.001$, ** $p<0.01$, * $p<0.05$.}
    \label{fig:ablation}
\end{figure*}

\begin{table}[t]
\centering
\begin{threeparttable} 
\caption{Comparison of All-session Experimental Results (Accuracy \% $\pm$ Std)}
\label{tab:all_session}
\renewcommand{\arraystretch}{1.1} 
\setlength{\tabcolsep}{5pt} 
\footnotesize

\begin{tabular}{lccc}
\toprule
\textbf{Method} & \textbf{SEED} & \textbf{SEED-IV} & \textbf{SEED-V} \\
\midrule
DCCA-AM \cite{liu2022identifying} & 84.04$\pm$7.35 & - & - \\
CoDF-Net \cite{gong2023codf} & 87.04$\pm$6.58 & - & - \\
MSBLS \cite{gong2024cross} & 90.16$\pm$6.40 & - & - \\
RHPRNet \cite{tang2024hierarchical} & - & 67.62$\pm$7.26 & 68.44$\pm$12.25 \\
CFDA-CSF \cite{Jimenez2024CFDA} & 90.47$\pm$5.08 & 80.63$\pm$6.14 & 84.89$\pm$11.30 \\
MMDA \cite{10819285} & \textbf{94.82}$\pm$\textbf{2.41} & 85.54$\pm$8.11 & 86.77$\pm$10.32 \\
\midrule 
\textbf{IDEAL (Ours)} & 94.31$\pm$4.55 & \textbf{86.79}$\pm$\textbf{4.50} & \textbf{88.13}$\pm$\textbf{9.40} \\
\bottomrule
\end{tabular}
\end{threeparttable}
\end{table}

\textbf{2) All-Session Results.} Table \ref{tab:all_session} summarizes the performance under the challenging All-Session protocol. On SEED, IDEAL achieves \textbf{94.31\%}, comparable to MMDA (94.82\%) and significantly higher than CFDA-CSF (90.47\%). On \textbf{SEED-IV}, our method achieves \textbf{86.79\%}, outperforming the second-best MMDA (85.54\%) by 1.25\%, with a strictly lower standard deviation (4.50\% vs. 8.11\%). On \textbf{SEED-V}, IDEAL attains \textbf{88.13\%}, establishing a new state-of-the-art by surpassing MMDA (86.77\%) and RHPRNet (68.44\%). All of these results confirm that IDEAL effectively handles the compounded domain shift introduced by mixing multi-session data.

\subsection{Ablation Studies}

We conducted a leave-one-component-out ablation study to quantify the contribution of each module. As illustrated in Figure \ref{fig:ablation}, we evaluated variants by removing: domain expansion (DE), contrastive loss ($\mathcal{L}_{\text{contrast}}$), ArcFace Loss ($\mathcal{L}_{\text{arc}}$), and hierarchical adversarial loss ($\mathcal{L}_{\text{domain}}$). Additionally, we replaced $\mathcal{L}_{\text{domain}}$ with a single-layer DANN to verify the necessity of the hierarchical design. 
\textbf{Key observations include:}
\begin{enumerate}
    \item \textbf{Criticality of Alignment.} Removing domain expansion or hierarchical adversarial loss ($\mathcal{L}_{\text{domain}}$) causes the most precipitous drop in accuracy across all datasets ($p<0.001$). This validates that addressing domain shift at both instance and feature levels is fundamental to the framework's success.
    \item \textbf{Synergy of Losses.} While removing $\mathcal{L}_{\text{arc}}$ or $\mathcal{L}_{\text{contrast}}$ individually yields moderate declines, their combined presence (as shown in the visualization section) is crucial for optimal class separability.
    \item \textbf{Hierarchical Superiority.} The multi-level adversarial strategy used in IDEAL consistently outperforms the single-level DANN variant, confirming that aligning features at varying abstraction levels is essential for multimodal signal adaptation.
\end{enumerate}

\subsection{\textit{Quantitative Analysis of Domain Expansion Strategy}}

We empirically evaluate the proposed instance-level expansion mechanism. In addition, we include the adaptive threshold screening experiments in the Supplementary Material.

\begin{figure}[t]
    \centering
    \includegraphics[width=0.9\linewidth]{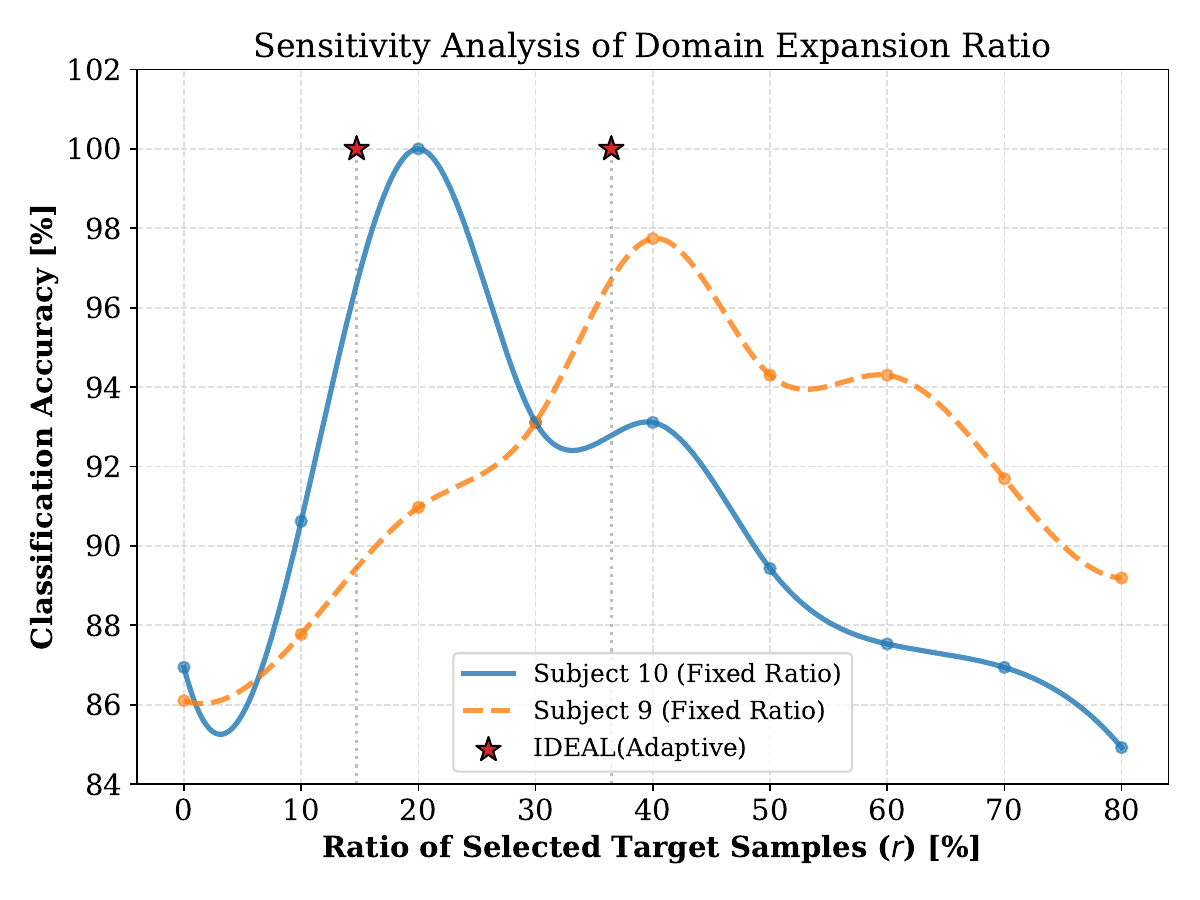}
    \caption{\textbf{Sensitivity analysis of the domain expansion ratio ($r$) on model performance.} The curves depict the accuracy of IDEAL on two representative subjects when manually fixing the screening ratio. Both subjects exhibit an inverted-U trend, confirming the trade-off between sample quantity and quality. The red stars indicate the performance of our proposed adaptive screening mechanism. IDEAL automatically identifies the optimal subject-specific ratio without manual tuning, achieving the peak accuracy of 100\% in both cases.}
    \label{fig:sensitivity}
\end{figure}

\textbf{1) Analysis of the screening Ratio.} 
Sensitivity analysis of the selection ratio (Figure~\ref{fig:sensitivity}) reveals a strict inverted-U relationship between the selection ratio \( r \) and the final classification accuracy, confirming the intrinsic quantity-quality trade-off formalized in Eq.~\eqref{eq:optimization}. Under-screening (\( r \to 0 \)): the mixing proportion \( \gamma = |C| / (|D_S| + |C|) \) approaches zero. Although the selected subset \( C \) strictly satisfies \( d(C, D_T) \ll d(D_S, D_T) \), the marginal tightening of the generalization bound becomes negligible due to insufficient transferred mass. Over-screening (\( r \to 1 \)): low-consensus noisy samples violate \( d(C, D_T) < d(D_S, D_T) \), directly breaking the prerequisite of \textbf{Proposition~1}, thus leading to severe negative transfer and abrupt performance drops. Notably, our proposed adaptive objective automatically converges to the subject-specific optimal ratio (e.g., 14.73\% vs. 36.46\% for different subjects, marked by red stars in Figure~3), achieving peak performance without any manual hyperparameter tuning.

\begin{figure}[t]
    \centering
    \includegraphics[width=0.9\linewidth]{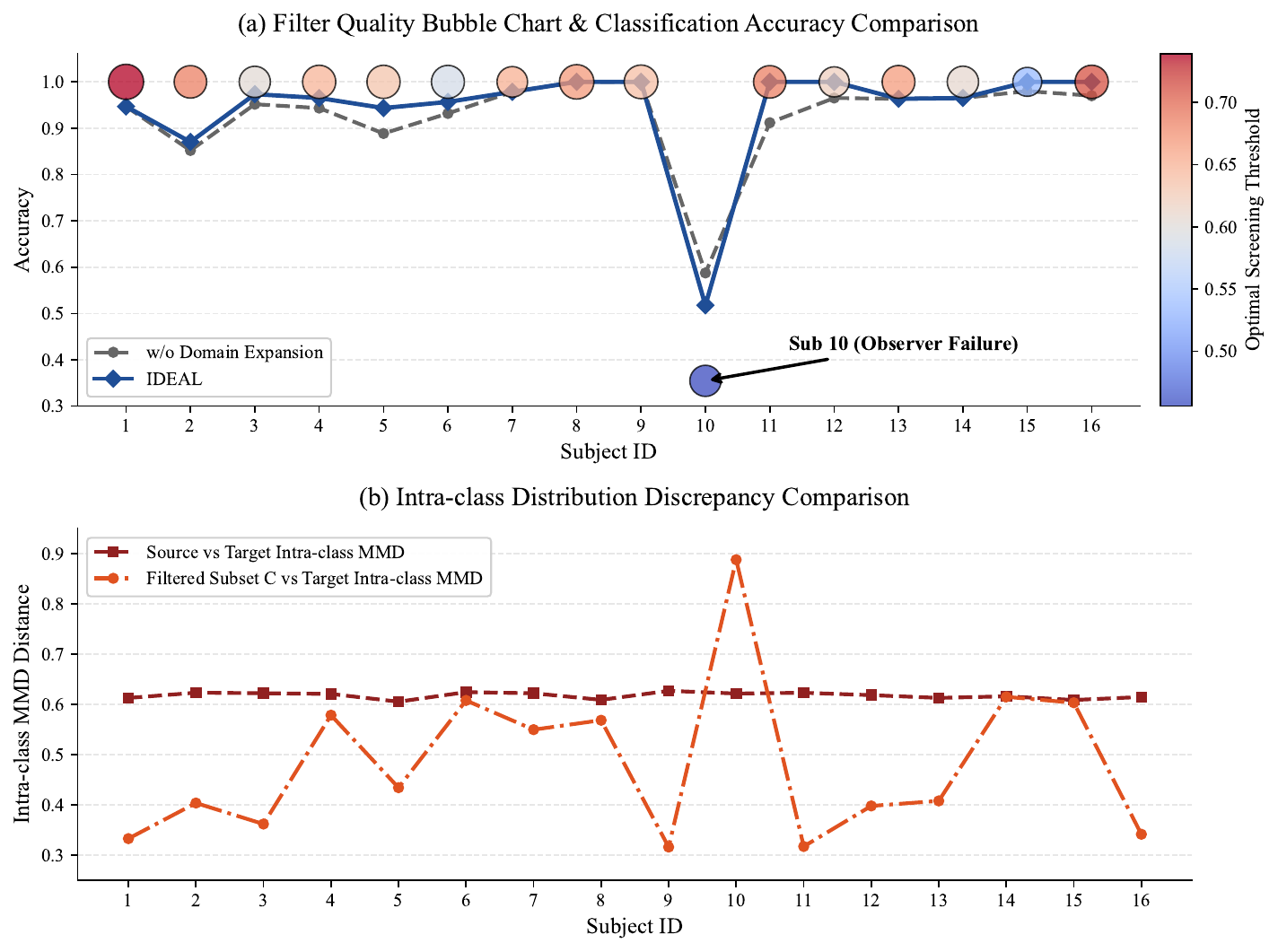}
    \caption{
\textbf{SEED-V (Session 3) instance-level domain expansion analysis across 16 subjects. }
(a) Bubble chart: bubble size represents selected count, color represents optimal threshold, y-axis represents pseudo-label accuracy. 
(b) Intra-class MMD: orange dash-dotted curve lies below red dashed, verifying \(d_{\mathcal{H}\Delta\mathcal{H}}(C, D_T) < d_{\mathcal{H}\Delta\mathcal{H}}(D_S, D_T)\). Subject~10 violates this condition.
}
    \label{fig:bubble}
\end{figure}

\textbf{2) \textbf{Mechanism Verification and Boundary Cases.}} 
We empirically validate \textbf{Condition~1} using class-aware MMD\cite{DDA} as the proxy for the $\mathcal{H}\Delta\mathcal{H}$-divergence in the \textbf{LOSO} experiments on SEED-V(Figure \ref{fig:bubble}). 
For the vast majority of subjects (e.g., Sub~1-9, 11-16), the screened subset $C$ strictly satisfies $d_{\mathcal{H}\Delta\mathcal{H}}(C, D_T) < d_{\mathcal{H}\Delta\mathcal{H}}(D_S, D_T)$. Consequently, the augmented domain $\mathcal{D}_S^*$ reduces the distributional gap, tightening the target generalization upper bound. This is quantitatively reflected in the high pseudo-label accuracy on $C$ and ultimately translated into substantial overall accuracy gains (Figure \ref{fig:bubble}(a)). 
However, Subject~10 serves as a controlled counterexample. Due to the presence of observer errors, multiple observers make mistakes simultaneously (though with low probability), causing the screened subset $C$ to violate the \textbf{Condition~1}. While this leads to marginal performance degradation for this specific case, extensive statistical analysis across all four datasets reveals that \textbf{over 90\% of subjects}—including the notoriously hard Subject~9 in SEED-IV with severe domain shift—successfully satisfy the screening condition and achieve positive transfer. This demonstrates the strong robustness of our instance-level expansion against physiological variability, with the failure rate confined to a statistically insignificant minority ($<10\%$, $p<0.001$ across all trials).

\begin{figure}[t]
    \centering
    \includegraphics[width=1\linewidth]{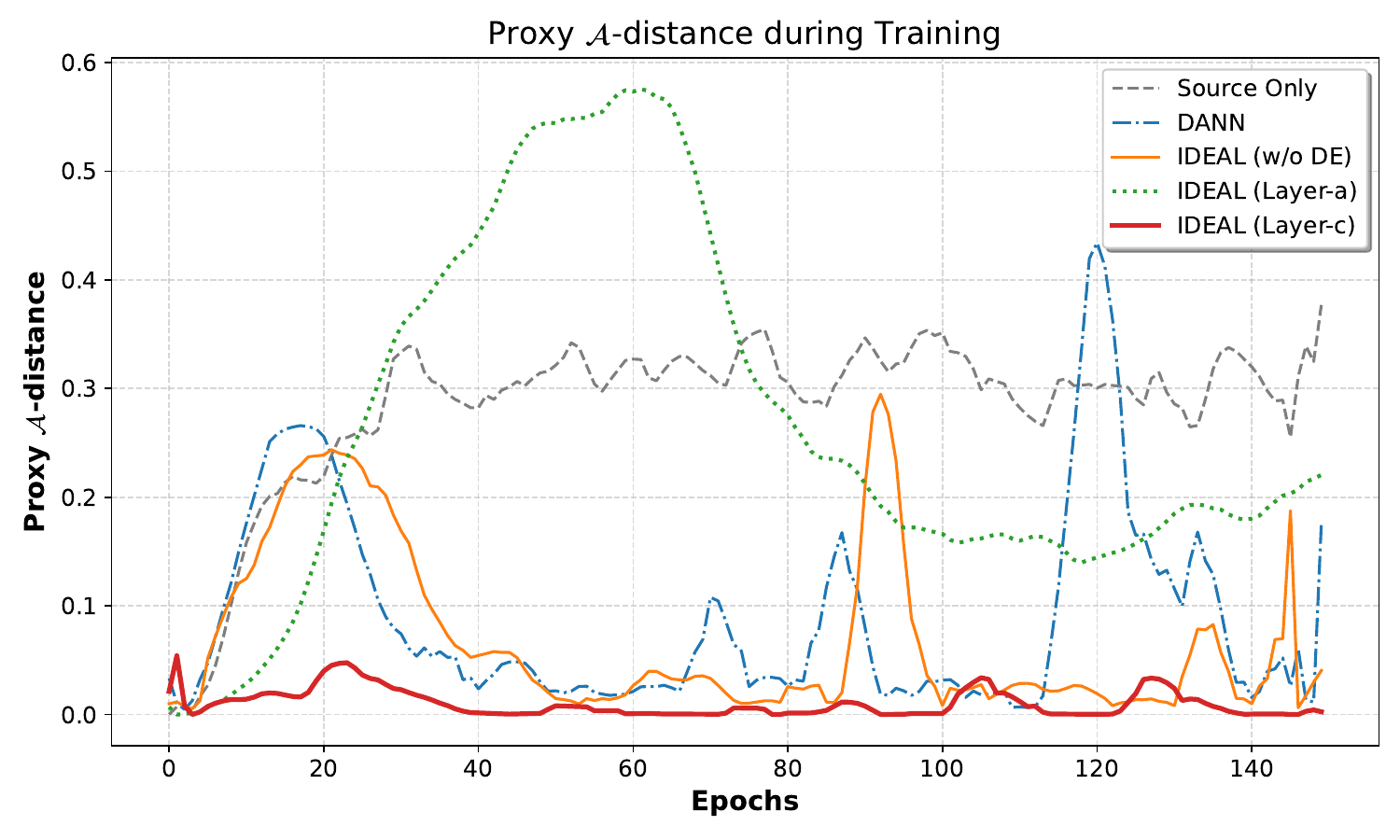}
    \caption{\small \textbf{Dynamic Evolution of Proxy $\mathcal{A}$-distance (PAD) during Training.} It quantifies the domain discrepancy between Source and Target domains across training epochs. \textbf{Source Only} exhibits increasing divergence due to overfitting. \textbf{DANN} reduces discrepancy but fluctuates significantly. In contrast, our \textbf{IDEAL (Layer-c)} achieves the lowest PAD, demonstrating near-perfect alignment. Notably, the gap between \textbf{Layer-a} and \textbf{Layer-c} verifies our hierarchical design: domain invariance is progressively learned from shallow to deep layers. The instability of \textbf{w/o DE} further confirms the stabilizing role of instance-level domain expansion.}
    \label{fig:pad_curve}
\end{figure}

\subsection{Convergence Analysis and Visualization}

\textbf{1) Dynamic Convergence Analysis (Proxy $\mathcal{A}$-distance).}
We monitor the Proxy $\mathcal{A}$-distance (PAD) \cite{ben2010theory} during training to assess domain alignment. PAD is defined as $d_A = 2(1 - 2\epsilon)$, where $\epsilon$ is the error of a domain classifier distinguishing source and target features; lower values indicate better alignment.

As shown in Figure~\ref{fig:pad_curve}, the Source Only baseline shows an increasing PAD trend, indicating that the model learns increasingly domain-specific features without adaptation. DANN reduces PAD relative to the baseline but fluctuates considerably, converging to $d_A \approx 0.17$, highlighting the challenge of global alignment in multimodal high-dimensional spaces.

IDEAL achieves more stable convergence. Notably, the PAD of deep features (Layer-c) is consistently lower than that of shallow features (Layer-a), confirming that domain invariance is progressively refined across abstraction levels. The variant without instance-level expansion (w/o DE) shows higher variance and slower convergence, validating our theoretical claim that instance-level expansion serves as distributional pre-conditioning for adversarial alignment. Ultimately, IDEAL (Layer-c) reaches near-zero PAD ($\approx 0.002$), demonstrating successful domain-invariant representation learning while retaining task-relevant information.

\begin{figure*}[t]
    \centering
    \includegraphics[width=1.0\linewidth]{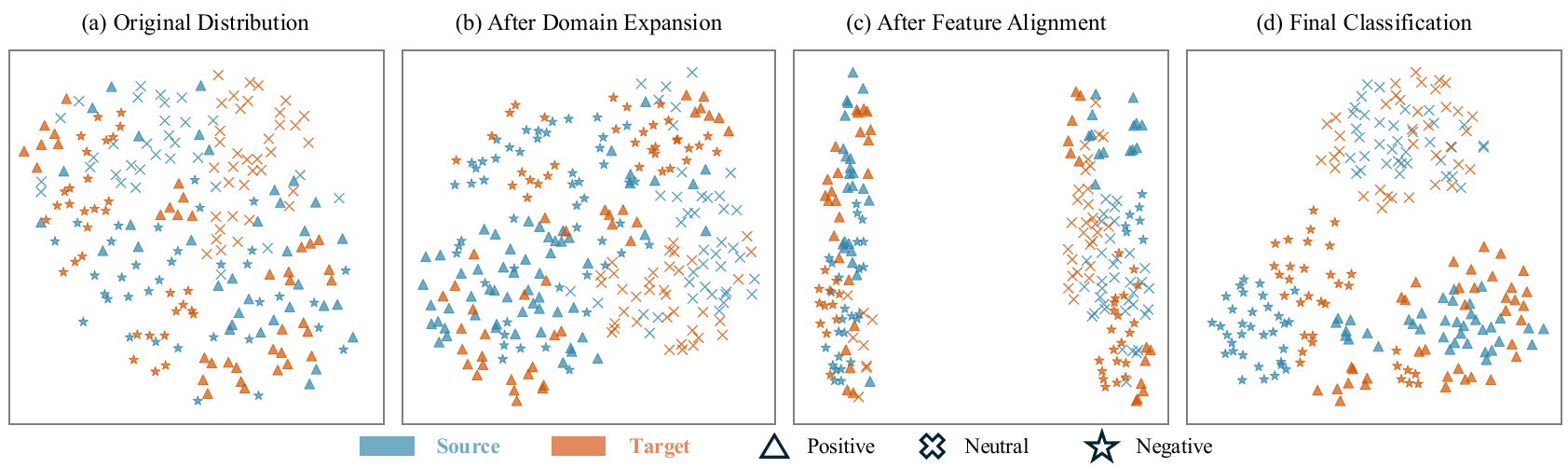}
    \caption{\small UMAP Visualization of Domain Adaptation Process. (a) Initial disjoint distributions of Source and Target domains. (b) Domain Expansion pulls distributions closer. (c) Hierarchical feature alignment further reduces discrepancy. (d) Final discriminative emotion clusters.}
    \label{fig:umap}
\end{figure*}

\textbf{2) Distribution Alignment.}
To intuitively verify the domain expansion mechanism, we visualized the feature distributions of Subject 10 (Session 3, SEED) using UMAP \cite{UMAP}, as shown in Figure \ref{fig:umap}. Figure \ref{fig:umap}(a) shows the raw input, where source and target domains are disjoint with significant distribution shift. Figure \ref{fig:umap}(b) demonstrates that after instance-level domain expansion, the domains converge significantly, increasing the overlap of shared semantic regions. Figure \ref{fig:umap}(c) reveals that after feature-level adaptation, the distributions are well-aligned. Figure \ref{fig:umap}(d) confirms that the final learned features form distinct, compact clusters for Positive, Neutral, and Negative emotions.

Furthermore, we provide additional visualization and quantitative analysis of the impact of contrastive loss and arcface loss on class boundaries in the Supplementary Material.

\subsection{Discussion}
\label{sec:discussion_exp}

The experimental results demonstrate that IDEAL consistently outperforms state-of-the-art methods across four benchmark datasets under two LOSO protocols. Several factors contribute to this superior performance. First, the instance-level domain expansion stage explicitly narrows the raw distributional gap, effectively preconditioning the feature-level alignment and reducing negative transfer—a claim supported by both theoretical bound tightening (\textbf{Proposition~1}) and empirical validation (Figure \ref{fig:bubble}). Second, the hierarchical adversarial module enforces domain invariance progressively across abstraction levels, which is particularly beneficial for physiological signals where emotional information is encoded at multiple scales. Third, the adaptive threshold selection eliminates manual tuning and automatically balances quantity-quality trade-offs according to subject-specific domain shifts.

Nevertheless, IDEAL has two primary limitations. The screening mechanism relies on consensus among base classifiers (RF, ET, CatBoost); if these observers lack diversity or accuracy—as observed in the rare failure case of Subject~10 (Figure \ref{fig:bubble})—the screened subset may violate the screening condition, leading to marginal performance degradation. Additionally, the two-stage architecture incurs higher offline computational cost (Stage~1 precomputation), though the deployable training time remains comparable to end-to-end methods. Future work will explore differentiable integration of instance selection via meta-learning, extend the framework to handle missing modalities through cross-modal imputation, and strategically weight multiple source domains to further enhance generalization in unconstrained real-world scenarios.

\section{Conclusion}
\label{sec:conclusion}
This paper proposes IDEAL, a unified framework for robust cross-subject emotion recognition with EEG and eye movement signals. By synergizing instance-level curriculum expansion and feature-level hierarchical adaptation, IDEAL effectively tackles the coupled challenges of individual-induced domain shift and cross-modal heterogeneity. 

The core contribution of this work is the establishment of a \emph{Curriculum-then-Alignment} philosophy. First, a multi-model collaborative screening mechanism functions as a curriculum filter, identifying high-confidence target samples to explicitly bridge the data-level distributional gap. Subsequently, a multimodal Transformer augmented with three-level adversarial constraints learns representations that are both domain-invariant and semantically discriminative. Extensive experiments on the SEED series datasets demonstrate that IDEAL significantly outperforms state-of-the-art baselines, validating its potential as a robust paradigm for generalized physiological computing. We believe this work not only advances the technical frontier of affective computing but also takes a significant step towards creating more empathetic and reliable brain-computer interfaces for real-world applications.

\bibliographystyle{IEEEtran} 
\bibliography{reference} 

\vfill

\end{document}